\documentclass[a4paper,11pt]{amsart}
\begin{document}

\hyphenation{gra-vi-ta-tio-nal re-fe-ren-ces Re-la-ti-vi-ty
ge-ne-ra-ted ma-the-ma-ti-cal ana-ly-ti-cal me-cha-nism
ex-pe-ri-men-tal ge-ne-ral ap-pro-xi-ma-tion po-pu-la-tion
sui-ta-ble re-fe-ren-ce}
\title[Gravitational waves from BBH-systems? A (doubly) vain quest]
{{\bf Gravitational waves from BBH-systems?
\\ A (doubly) vain quest}}
\author{Angelo Loinger}
\date{}
\address{Dipartimento di Fisica, Universit\`a di Milano, Via
Celoria, 16 - 20133 Milano (Italy)}
\email{angelo.loinger@mi.infn.it}
\date{}
\thanks{To be published on \emph{Spacetime \& Substance}}

\begin{abstract}
 The theoretical reasons at the root of LIGO's experimental
 failure in searching gravitational waves (GW's) from binary black
 hole (BBH) inspirals.
\end{abstract}

\maketitle

\noindent \emph{\textbf{Summary}} -- \textbf{1}. A recent LIGO's
search for GW's. -- \textbf{2}. BH's and BBH's are fictive
objects. -- \textbf{3}. Experimental confirmation: no BH's and no
BBH's have ever been \emph{really} observed. -- \textbf{4}. The
GW's according to \emph{linear} approximation of GR: ghost
undulations. -- \textbf{5}. There exists \emph{no} physical
``mechanism" of production of GW's in the \emph{exact} GR: they
are phantom entities; and the experience confirms. -- \textbf{6}.
Conclusion. -- Appendices A, B, C, D. \vskip1.20cm

\noindent {\bf 1.}-- The \emph{\textbf{abstract}} of a recent
paper by LIGO scientific collaboration \cite{1} runs as follows:
``We report on a search for gravitational waves from binary black
hole inspirals in the data from the second $[2003]$ science run of
the LIGO interferometers. The search focused on binary systems
with component masses between $3$ and $20$ solar masses. Optimally
oriented binaries with distances up to $1$Mpc could be detected
with efficiency of at least $90\%$. We found no events that could
be identified as gravitational waves in the $365.6$ hours of data
that we searched."

\par This is not the first LIGO's failure in the quest of a source
of gravitational waves (GW's); however, the candid optimism of the
authors remains (seemingly) intact. At p.13 of \cite{1} we read:
``$\ldots$ the characteristics of the BBH population (such as
spatial, mass and spin distributions) are not known, since no BBH
systems have ever been observed." And at p.16, in
\textbf{\emph{Conclusions and future prospects}}: ``This search,
even though similar in some ways to the binary neutron star
$[$BNS$]$ inspiral search $[$a previous complete failure!$]$, has
some significant differences and presents unique challenges. There
were no events that could be identified as gravitational waves. --
The fact that the performance and sensitivity of the LIGO
interferometers is improving and the frequency sensitivity band is
being extended to lower frequencies makes us hopeful that the
first detection of gravitational waves from the inspiral phase of
binary black hole coalescences may happen in the near future. In
the absence of a detection, astrophysically interesting results
can be expected by LIGO very soon." (What kind of results?) --

\vskip0.80cm
\noindent {\bf 2.}-- Unfortunately for LIGO's scholars, their
hopes are doomed to be frustrated -- and for several reasons.

\par First of all, the general form of solution of Schwarzschild
problem is given, in spherical polar co-ordinates, by the
following $\textrm{d}s^2$ \cite{2}:

\begin{eqnarray} \label{eq:one}
\qquad \qquad \textrm{d}s^2 = \left(1-\frac{2m}{f(r)}\right)
c^{2}\textrm{d}t^2 - \left(1-\frac{2m}{f(r)}\right)^{-1}
[\textrm{d}f(r)]^{2} -  \nonumber\\ -[f(r)]^{2}
\left(\textrm{d}\vartheta^{2}+\sin^{2}\vartheta
\textrm{d}\varphi^2\right) \quad{,}
\end{eqnarray}

where: $m \equiv GM/c^{2}$; $M$ is the mass of the concerned
material point (at rest); $f(r)$ is \emph{any} regular function of
the radial co-ordinate $r$. Remark that no physical result depends
on the choice of $f(r)$. If we put

\begin{equation} \label{eq:two}
f(r) \equiv \left[ r^{3} + (2m)^{3} \right]^{1/3} \quad{,}
\end{equation}

we obtain the \textbf{\emph{original}} form of solution given by
Schwarzschild \cite{3}, which is \emph{maximally extended} since
holds for $r>0$. Putting

\begin{equation} \label{eq:three}
f(r) \equiv r \quad{,}
\end{equation}

we have the \textbf{\emph{standard}} (by Hilbert-Droste-Weyl) form
(\emph{erroneously} called ``by Schwarzschild"), which holds
\emph{only} for $r>2m$, as it was repeatedly emphasized by
\emph{all} the Fathers of Relativity; remark that the exterior
part $r>2m$ of form (\ref{eq:three}) is diffeomorphic to form
(\ref{eq:two}). The notion of black hole (BH) is a senseless
artefact based on an unphysical interpretation of the singular
locus $r=2m$ and of the region $r<2m$ \cite{4}. Moreover, an
appropriate computation shows that \emph{a continued gravitational
collapse of a massive celestial body ends in a} \textbf{\emph{mass
point}}\emph{, not in a BH} \cite{4}.

\par So far as the binary black hole (BBH) systems are concerned,
it is sufficient to remember a sentence written by McVittie many
years ago, in which he emphasized that, even if one admits the
existence of BH's (on which he was frankly sceptical), ``there is
no way of asserting through some analogy with Newtonian
gravitational theory that a black hole could be a component of a
close binary system or that two black holes could collide. An
existence theorem would first be needed to show that Einstein's
field equations contained solutions which described such
configurations." \cite{5}.

\vskip1.20cm
\noindent {\bf 3.}-- From the \emph{observational} standpoint, the
``observed" BH's are only large, or enormously large, masses
concentrated in very small volumes; indeed, no observation has
ever detected any whatever characteristic property of the
hypothetical BH's, e.g. the existence of a ``horizon". On this
matter we owe to Wolfgang Kundt many interesting remarks \cite{6},
for instance: ``How about stellar-mass holes? $[\ldots]$ Over 45
\emph{black hole candidates} have been proposed during the past 30
years from the class of binary X-ray sources $[\ldots]$. To me,
all of them look like neutron stars surrounded by massive
($\approx 5$ solar masses) accretion disks, because of their often
hard spectra (up to the $\gamma$-ray range), highly structured,
$[\ldots]$ and because of their indistinguishable further
properties, as a class, from all the established neutron-star
binaries $[\ldots]$. -- And the postulated \emph{supermassive
black holes} at the centers of (all the active) galaxies?
$[\ldots]$ Active galactic nuclei may owe their extreme properties
to those of their central disks." (See pp.101 and 102 of
\cite{6}).

\par And ``no BBH systems have ever been observed." \cite{1}.

\vskip1.20cm
\noindent {\bf 4.}-- In the \emph{exact} (\emph{non}-linearized)
formulation of GR the GW's are only mathematical undulations fully
destitute of physical reality, as it was first proved in 1917 by
Tullio Levi-Civita \cite{7} -- and subsequently with different
demonstrations by, in particular, Scheidegger \cite{8}, Infeld and
Plebanski \cite{9}, and the present writer \cite{10}.

\par How about GW's in the \emph{linear} approximation of GR?
Strictly speaking, a detailed investigation is superfluous,
because \emph{ubi maiora minora cessant}. However, since all
experimentalists and many theoreticians have inadequate ideas on
this matter, and are fond of the linear version of GR \cite{11},
it is suitable to re-examine the question.

\par In 1923 Eddington published a limpid paper in which, with
utilization of previous results by Einstein and Weyl, the
conceptual bases of the \textbf{\emph{linear}} theory of the GW's
were definitely settled \cite{12}.

\par As it is well known, the crucial postulate of the linear
approximation of GR is the following: one sets approximately:

\begin{equation} \label{eq:four}
g_{jk} \approx \eta_{jk} + h_{jk} \quad, \qquad((j,k=0,1,2,3))
\quad,
\end{equation}

where $\eta_{jk}$ is the customary Minkowskian tensor and the
$h_{jk}$'s are small deviations, that in our case represent the
passage of GW's. \emph{Remark that, quite generally,} $h_{jk}$
\emph{is a tensor} \textbf{\emph{only}} \emph{under Lorentz
transformations of co-ordinates}. Eddington  considers \emph{in
primis} -- in a \textbf{\emph{generic}} Cartesian frame -- the
emblematic instance of plane waves proceeding with a velocity $V$
(not fixed \emph{a priori}!) in the (negative) direction of the
axis of $x$, so that the $h_{jk}$'s are periodic functions of
$(x+Vt)$ only. He finds that the $h_{jk}$'s of the undulatory
disturbance can be separated into three \emph{independent} sets:
\\

\qquad Transverse-transverse: $h_{22}$, $h_{33}$, $h_{23}$; \quad
($h_{22}=-h_{33})$ ; \

\qquad Longitudinal-transverse: $h_{12}$, $h_{13}$, $h_{02}$,
$h_{03}$ ; \

\qquad Longitudinal-longitudinal: $h_{11}$, $h_{01}$, $h_{00}$.
\\

\par One puts: $h_{+}\equiv \{h_{22}=-h_{33}\}$, and $h_{\times}\equiv
\{h_{23}=h_{32}\}$; these \emph{traceless} types can be
transformed into each other with a space rotation of the
co-ordinates $x^{2}, x^{2}$ around the $(x\equiv x^{1})$-axis by
an angle of $\pi/4$ radians.

\par Now, the Author proves that the transverse-trasverse waves
are propagated with the velocity $c$ of light \emph{in vacuo},
whereas the other waves are propagated with an arbitrary velocity,
e.g. with the speed of thought. The curvature tensor depends only
on the transverse-transverse waves; the remaining waves are only
analytical fictions, which disappear with a suitable
(\emph{infinitesimal}) change of co-ordinates. Accordingly, it
seems that the transverse-transverse waves are endowed with a a
physical reality. I affirm that this is a mere illusion. The
simple proof runs as follows: \emph{i}) the linear version of GR
is only an \textbf{\emph{approximate}} formulation; \emph{ii}) in
GR there is \textbf{\emph{no}} class of physically privileged
reference frames; moreover, for the hypothetical GW's -- solutions
of Einstein equations $R_{jk}=0$ -- \textbf{\emph{all}}
co-ordinate systems are physically suitable and appropriate;
\emph{iii}) a \textbf{\emph{finite}} transformation of
\textbf{\emph{general}} co-ordinates can reduce to
\textbf{\emph{zero}} the transverse-transverse undulations, which
have a tensor character under Lorentz transformations
\textbf{\emph{only}}. \emph{An immediate corollary: the watts of
the celebrated mass-quadrupole formula are only ghost watts for
the} \textbf{\emph{exact}} \emph{GR}. (According to some authors,
the validity of this formula goes \emph{beyond} the linear
approximation. Well, even in this case its physical value would
remain inexistent owing to its derivation from the gravitational
energy-momentum \emph{pseudo} tensor, which can be always reduced
to zero by means of an appropriate change of reference system.)

\vskip1.20cm
\noindent {\bf 5.}-- In 1944 Hermann Weyl published a profound
memoir on the linear version of GR \cite{13}. He emphasized that
in reality the gravitational field of this version is a
``powerless shadow"  because it exerts \textbf{\emph{no}} force on
matter. Indeed, ``the gravitational force arises only when one
continues the approximation beyond the linear stage $[\ldots]$": a
fundamental result  of Einstein-Infeld-Hoffmann method (see e.g.
\cite{9}). An immediate consequence: the customary linearized
computations of the action of a GW on matter (for instance, on a
resonant bar or on a Michelson interferometer) are fully destitute
of a physical sense.

\par If we continue the approximation beyond the linear stage, we
find that the \emph{radiation terms} of the gravitational field
can be always \textbf{\emph{destroyed}} by suitable co-ordinate
transformations: the GW's are, quite generally, analytical
phantoms, mere products of special choices of the reference frame
(\cite{8}, \cite{9}). This result can be also demonstrated
independently of any perturbative treatment; stringently, it can
be proved that \emph{there exists \textbf{no} ``mechanism" of
production of GW's} \cite{10}: in the last analysis, this depends
on the fact that in GR the motions of the particles result in
principle from Einstein \textbf{\emph{field}} equations -- as it
was first demonstrated in the fundamental memoir of 1927 by
Einstein and Grommer.

\par (There are several proofs of the real non-existence of the
above ``mechanism". For instance: it is very easy to prove that
the trajectories of the bodies of a physical system, which
interact only gravitationally -- as e.g. the bodies of the solar
system -- , are \emph{geodesic} lines, and therefore any emission
of GW's is clearly impossible. This result can be extended to the
cases in which there are also \emph{non}-gravitational
interactions.)

\vskip1.20cm
\noindent {\bf 6.}-- \emph{\textbf{Conclusion.}} The black holes
are unreal objects and the theory of the gravitational waves based
on the \emph{linear} approximation of General Relativity is not
tenable from a rigorous standpoint.
\par More generally and definitively, the \emph{exact} formulation
of General Relativity tells us that the gravitational waves are
only phantasmic entities -- see particularly \cite{7}, \cite{8},
\cite{9}, \cite{10}.
\par It is time that unfounded ideas on General Relativity start
getting abandoned.

\small \vskip0.5cm
\par\hfill {\emph{``Betr\"ubt euch nicht, ihr guten Seelen!}
  \par\hfill \emph{Denn wer nicht fehlt, wei\ss{} wohl wenn andre fehlen $\ldots$"}
  \par\hfill J.W. v. Goethe

\normalsize

\vskip0.80cm
\noindent {\bf \small{APPENDIX A}} \par \vskip0.10cm According to
an erroneous conjecture, any GW generates a given gravitational
field. Now, the stress-energy-momentum tensor of a hypothetical GW
is a \emph{pseudo} tensor, which can be reduced to zero with a
change of reference system. Only the matter tensor $T_{jk}$
generates Einsteinian gravitational fields, just as only the mass
density $\mu$ generates Newtonian gravitational fields. Under this
respect, the non-linearity of Einstein field equations is not
important -- and it is remarkable that the supporters of the above
surmise affirm its validity even for the linear approximation of
GR (see e.g. at p.391 of the book by Landau \emph{et al.}
\cite{12}), neglecting further the fact that $h_{jk}$ is a true
tensor under Lorentz transformations only.

\vskip0.80cm
\noindent {\bf \small{APPENDIX B}}
\par \vskip0.10cm
Point \emph{ii}) of the last paragraph of sect.\textbf{4.} is also
important for a plain proof of the physical inadequacy of any
\emph{exact} theory of GW's, for instance of the theory of the
plane waves by Bondi \emph{et al.} \cite{14}. Indeed, convenient
transformations of general co-ordinates can destroy the undulatory
character of the wavy disturbances, in particular of the above
plane GW's. (Another consequence of point \emph{ii}) is the
following: \emph{no} fundamental velocity exists in \emph{general}
relativity -- contrary to a current belief.) Further, also the
stress-energy-momentum \emph{pseudo} tensor of Bondi's plane waves
can be obviously reduced to zero through an appropriate change of
reference frame.

\par Finally, it is true that there is in general the possibility
(see e.g. Pirani in \cite{14}) of building exact wavy solutions of
field equations for empty spacetime ($R_{jk}=0$), which have a
curvature tensor different from zero, but this is only a
mathematical possibility, for the simple reason that (in the
\emph{exact} GR) there is no physical ``mechanism" apt to generate
GW's. (And the above wavy solutions do not possess a \emph{true}
stress-energy-momentum tensor, but only a \emph{pseudo} one.)

\vskip0.80cm
\noindent {\bf \small{APPENDIX C}}
\par \vskip0.10cm
The first sentence of paper \cite{1} is: ``The Laser
Interferometric Gravitational Wave Observatory (LIGO) $[\ldots]$
consists of three Fabry-Perot-Michelson interferometers, which are
sensitive to the minute changes that would be induced in the
relative lengths of their orthogonal arms by a passing
gravitational wave."

\par The above ``minute changes" represent a \emph{mechanical}
effect produced by the passage of a (hypothesized) GW. Now, such a
wave would be capable also of production of \emph{electromagnetic}
effects, since it would interact with the light beams in the
Fabry-Perot-Michelson cavities.This interaction has been the
object of an accurate study by Cooperstock and Faraoni, see their
paper quoted in \cite{12}. These authors solved the problem in the
\emph{\textbf{linear}} approximation of GR, taking properly into
account the boundary conditions for the light beams. They computed
the phase shift, the light deflection and the rotation of the
polarization axis induced by GW's. Their results hold also for
detectors which are \emph{large} in comparison with the
wavelengths of the GW's. The great majority of the concerned
astrophysicists, in particular LIGO's members, do not mention the
electromagnetic effects of the GW's in the interferometric
detectors. Why? Are they afraid of an \emph{embarras de richesse}?

\vskip0.80cm
\noindent {\bf \small{APPENDIX D}}
\par \vskip0.10cm
The conclusion of the paper by Hough \emph{et al.} quoted in
\cite{11} is astonishing. They write: ``$\ldots$ the 1993 Nobel
Prize in Physics was awarded to Hulse and Taylor for their
experimental observations and subsequent interpretations
$[$\emph{which are based essentially on the}
\textbf{\emph{linear}} \emph{approximation of GR}$]$ of the
evolution of the orbit of the binary pulsar PSR1913+16, the decay
of the binary orbit being consistent with angular momentum and
energy being carried away from this system by gravitational waves.
Thus it is now universally accepted that gravitational waves must
exist unless there is something seriously wrong with General
Relativity."

\par This is quite illogical: the authors believe that the
interpretations of Hulse and Taylor are fully adequate; now, such
interpretations rest on the linearized version of General
Relativity, in which Hough \emph{et al.} repose a complete trust
$\ldots$ -- In reality, the above interpretations are
inappropriate, as it can be proved \cite{15}.

\par The experimental failures in detecting the GW's are a proof
of the experimental adequacy of General Relativity -- if
\emph{rightly} understood.
\\


\begin{thebibliography}{99}

\bibitem{1}
B. Abbott \emph{et alii} (several hundreds of scholars),
\emph{arXiv:gr-qc/0509129 v1} (September 30th, 2005).

\bibitem{2}
A. S. Eddington, \emph{The Mathematical Theory of Relativity},
Second Edition (Cambridge University Press, Cambridge) 1960, p.94.

\bibitem{3}
K. Schwarzschild, \emph{Berl. Ber.}, (1916) 189; for an English
version by S. Antoci and A. Loinger see
\emph{arXiv:physics/9905030} (May 12th, 1999).

\bibitem{4}
See e.g. A. Loinger and T. Marsico,
\emph{arXiv:physics/0512237 v1} (December 24th, 2005)-- in course
of publication on \emph{Spacetime \& Substance}; and references
therein.

\bibitem{5}
G.C. McVittie, \emph{The Observatory}, \textbf{78} (1978) 272.

\bibitem{6}
W. Kundt, \emph{Astrophysics -- A Primer} (Springer-Verlag,
Berlin, \emph{etc.}) 2001; see in particular pp. 37, 81, 101, 102,
131, 132.

\bibitem{7}
T. Levi-Civita, \emph{Rend. Acc. Lincei}, \textbf{26} (1917) 381;
an English translation by S. Antoci and A. Loinger in
\emph{arXiv:physics/9906004} (June 2nd, 1999). See also H.A.
Lorentz, \emph{Amst. Versl.}, \textbf{25} (1916) 468, and T.
Levi-Civita, \emph{Rend. Acc. Lincei}, \textbf{11} (s.6a) (1930) 3
and 113.

\bibitem{8}
A.E. Scheidegger, \emph{Revs. Mod. Phys.}, \textbf{25} (1953) 451.

\bibitem{9}
L. Infeld and J. Plebanski, \emph{Motion and relativity},
(Pergamon Press, Oxford,  \emph{etc.}) 1960, see in particular
Chapt.VI.

\bibitem{10}
Cf. e.g. A. Loinger, \emph{arXiv:physics/0504018} (April 3rd,
2005) -- in course of publication in \emph{Spacetime and
Substance}; \emph{\textbf{and references therein}}.

\bibitem{11}
See e.g.: J. Hough, S. Rowan, B.S. Sathyaprakash,
\emph{arXiv:gr-qc/0501007} (January 4th, 2005); J. Hough, \'E.
\'E. Flanagan and S.A. Hughes, \emph{arXiv:gr-qc/0501041} (January
12th, 2005)

\bibitem{12}
A. S. Eddington, \emph{Proc. Roy. Soc.}, \textbf{102A} (1923) 268;
see also: Idem,  \emph{The Mathematical Theory of Relativity},
Second Edition (Cambridge University Press, Cambridge) 1960, sect.
\textbf{57} and Note \textbf{7} (p.246). For comprehensive
\emph{formal} developments on the linearized theory of the GW's
see e.g. L. Landau et E. Lifchitz, \emph{Th\'eorie du Champ}
(\'Editions Mir, Moscou) 1966, sects.\textbf{101} and
\textbf{104}. See further: F.I. Cooperstock and V. Faraoni,
\emph{Class.Quantum Grav.}, \textbf{10} (1993) 1189.

\bibitem{13}
H. Weyl, \emph{Amer. J. Math.}, \textbf{66} (1944) 591. -- In the
Appendix of a paper of mine -- \emph{Spacetime and Substance},
vol.\textbf{5}, No. 2(22), 2004, pp.53-56 -- I have reproduced the
``Introduction and Summary" and the sects. \textbf{1}, \textbf{2}
of Weyl's memoir, which are particularly relevant to present
theme.

\bibitem{14}
H. Bondi, F.A.E. Pirani and L. Robinson \emph{Proc. Roy. Soc.},
\textbf{A 251} (1959) 519. See also F.A.E. Pirani, \emph{Phys.
Rev.}, \textbf{105} (1957) 1089.

\bibitem{15}
A. Loinger, \emph{Spacetime \& Substance}, Vol.\textbf{6} (2005),
No.1(26), pp28-30. Also in \emph{arXiv:physics/0502089 v1}
(February 16th, 2005).


\end{thebibliography}
\end{document}